\documentstyle[prl,twocolumn,aps,graphicx]{revtex}
\begin{document}
\bibliographystyle{prsty}
\draft

{\bf K\"onig et al. Reply:}
In a recent Letter \cite{us} we developed a theory of carrier-induced
ferromagnetism in diluted magnetic semiconductors.
We analyzed the elementary spin excitations at low temperatures, where
spin waves can be approximated as non-interacting Bose particles 
(``independent spin-wave theory'').
In addition, we proposed a simple ad hoc ``self-consistent spin-wave'' 
approximation for higher temperatures in order to demonstrate the increasing 
inadequacy of mean-field theory critical temperature estimates at high carrier
densities.

Following Ref.~\cite{callen}, Yang {\it et al.} \cite{yang} use an 
equation-of-motion approach under the Tyablikov decoupling scheme to arrive at 
an alternative ``self-consistent spin-wave'' theory.
As we will show now, it is straightforward to rederive this scheme 
within our formulation, and thereby provide a clearer physical picture of the 
nature of their approximation.

The independent spin-wave theory describes small amplitude collective 
fluctuations of the magnetization, spin waves with dispersion $\Omega_p$.
It predicts for the thermal average of 
the impurity-spin density
\begin{equation}
   \langle S^z \rangle = {1\over V} \sum_{|\vec p| < p_c} \left\{
	S - n(\Omega_p) \right\} \, .
\label{sw}
\end{equation}
The Bose function $n(x)$ reflects the fact that the spin waves are 
approximated by independent Bose particles.
It is known \cite{dyson} that this equation yields the correct prefactor of the
characteristic $T^{3/2}$-law.

A self-consistent spin-wave theory provides an approximate theory of large 
amplitude magnetization fluctuations, taking into account, for example,
spin-quantization constraints on boson occupation numbers.

Weiss mean-field theory, which is expected to be accurate for a model with 
static long-range interactions, neglects correlation and, hence, spin-wave 
dispersion, but treats the problem self consistently.
Then the constraint on the number of spin waves per impurity spin ($\le 2S$) 
leads to $\langle S^z \rangle = c S B_S(\beta S \Omega)$ or, equivalently,
\begin{equation}
   \langle S^z \rangle = c \left\{
	S - n(\Omega) + (2S+1) n[(2S+1)\Omega]\right\}
\label{mf}
\end{equation}
where $B_S(x)$ is the Brillouin function, and $\Omega$ denotes 
the energy of an uncorrelated spin flip, which is independent of momentum 
$\vec p$.
The mean-field theory is closed by specifying the dependence of $\Omega$ on
$\langle S^z\rangle$ (it is $\Omega = J_{pd} n^*$ where $n^*$ is the 
free-carrier spin density).
The second Bose function in Eq.~(\ref{mf}) is the correction term from spin
kinematics and rules out unphysical states.
Due to the neglect of correlation, Weiss mean-field theory can substantially 
overestimate the critical temperature \cite{schlie}.
It fails to describe the low-temperature magnetization as well.

A self-consistent spin-wave scheme ideally (i) reduces to the independent 
spin-wave theory at low temperatures, (ii) simplifies to Eq.~(\ref{mf}) in the 
Ising limit, $\Omega_p \rightarrow \Omega$, and (iii) yields a second-order 
phase transition by allowing for the trivial solution $\langle S^z \rangle =0$.

At low temperatures the correction term in Eq.~(\ref{mf}) is negligible and
Eq.~(\ref{mf}) reduces to Eq.~(\ref{sw}) if $\Omega$ is replaced by an
effective energy which accounts for the spin-wave dispersion $\Omega_p$,
\begin{equation}
   n(\Omega) \equiv {1\over cV} \sum_{|\vec p| < p_c} n(\Omega_p) 
\label{cond}
\end{equation}
($=\Phi$ in the notation of Ref.~\cite{yang}). 
In Ref.~\cite{us} we obtained a self-consistent spin-wave scheme by using the 
replacement $\Omega \rightarrow \Omega_p$ in Eq.~(\ref{mf}) and averaging over 
momenta, $(1/cV)\sum_{|\vec p|<p_c} \ldots$.
This is equivalent to restrict the number of spin waves {\it per wave vector} 
$\vec p$ and turns out to be too restrictive at low temperatures, as we 
pointed out in Ref.~\cite{us}.
Nevertheless, requirements (ii) and (iii) are still satisfied and the 
prefactor of the $T^{3/2}$-law is only slightly changed.
This approximation would completely fail in reduced-dimension systems as 
emphasized in Ref.~\cite{yang} but these are not of interest here \cite{lee}.

Equation~(3) of Ref.~\cite{yang} is identical to Eq.~(\ref{mf}) with $\Omega$ 
determined by Eq.~(\ref{cond}).
Now all the requirements (i), (ii), and (iii) are satisfied.
The scheme proposed by Yang {\it et al.} is, therefore, equivalent to Weiss 
mean-field theory but with an effective (constant) spin-flip energy $\Omega$ 
which accounts for correlation.

Finally, we note that in both self-consistent schemes, the one in 
Ref.~\cite{us} and the one proposed in Ref.~\cite{yang}, not only does the 
magnetization of the impurity spins vanish at the critical temperature but,
by construction, the polarization of the free carriers goes to zero as well.
This is not generally the correct physical picture.
For example, in the limit of low carrier densities, the local band 
polarization of the free carriers remains finite even above 
$T_c$ \cite{schlie} and neither scheme gives reliable $T_c$ estimates.

\end{document}